\documentclass[11pt]{article}
\usepackage{amsmath}
\usepackage{caption}
\usepackage{multirow}
\usepackage{hyperref}
\usepackage[table]{xcolor}
\usepackage{graphicx}
\usepackage{amssymb}
\usepackage[latin1]{inputenc}
\usepackage{listings}
\usepackage{fancyvrb}
\usepackage{amsmath, amssymb, epsfig}
\usepackage{amsmath}
\usepackage{amsthm}
\usepackage{amssymb}
\usepackage{bm}
\usepackage{mathrsfs}
\usepackage{wasysym}

\usepackage{fullpage}

\theoremstyle{plain}

\theoremstyle{definition}

\theoremstyle{remark}

\numberwithin{equation}{section}

\def \< {\langle}
\def \> {\rangle}
\def \^ {\widehat}

\title{Statistical Learning for Best Practices in Tattoo Removal} 
\author{Richard P. Yim$^1$\thanks{Correspondence: richyim555@g.ucla.edu}, Jamie Haddock$^1$, Deanna Needell$^1$}
\date{%
    Affiliations:
    $^1$Department of Mathematics, University of California, Los Angeles%
}

\begin{document}
\maketitle \begin{abstract} \noindent The causes behind complications in laser-assisted
    tattoo removal are currently not well understood, and in the literature
    relating to tattoo removal the emphasis on removal treatment is on removal technologies and tools, not best
    parameters involved in the treatment process. 
    Additionally, the very challenge of determining best practices is difficult given the complexity of
    interactions between factors that may correlate to these complications. In this paper we apply a battery of classical 
    statistical methods and techniques
    to identify features that may be closely correlated to causes of complication during the tattoo removal
    process, and report quantitative evidence for potential best practices. We develop elementary statistical
    descriptions of tattoo data collected by the largest gang rehabilitation and reentry organization in the world,
    Homeboy Industries; perform parametric and nonparametric tests of significance; and finally, 
    produce a statistical model explaining treatment parameter interactions, 
    as well as develop a ranking system for treatment parameters utilizing bootstrapping and gradient boosting.
    \vspace{10pt}

    \noindent
\end{abstract}

\section{Introduction}

Current best treatment practices for laser-assisted tattoo removal are 
ad hoc at best, and the interactions between parameters involved in the tattoo removal
treatment process are not well understood. For instance, it is believed that patient demographic factors may potentially
affect treatment outcomes (e.g., factors such as patient age and gender), and that these factors 
are a source of variation in physiological processes between individuals (i.e., skin reactivity and healing time 
vary demographically). Additionally, much of the research surrounding tattoo removal is focused on the clinical
trial setting with an emphasis on discovering new innovations for laser-assisted tattoo removal  
\cite{baumler2019laser}. While there is considerable debate on laser pulse duration standards, it still remains that there
is no precise and standard system for best practices that has been shown to be both optimally effective and safe
\cite{kurniadi2020laser}. A reason for the lack of understanding of safe treatment practices is a general lack of 
records of treament parameters causing complications \cite{ho2015laser}.

For our study we explore data recorded by Homeboy Industries, the largest gang rehabilitation
program in the world, which offers free tattoo removal treatment services for former gang members seeking to remove gang-related tattoos.
We utilize various statistical methods and demonstrate some results of inference relating to laser-assisted tattoo
removal treatment procedures performed in a practical clinical setting at Homeboy Industries in Los Angeles. 

In our data there are four types of complications that are associated with tattoo removals: hyperpigmentation and hypopigmentation,
or increased and decreased skin pigmentation, respectively; scarring, visible tissue regrowth; and keloids, scarring with
excessive skin overgrowth in the treated area.
We emphasize that while the complication rate among the
studied sample is low, the main concern for all 
tattoo-removal practitioners is a commitment to doing as little harm
as possible. For this
reason, our primary concern with the available data is to simply understand complication occurrence and nonoccurence in
general, grouping all
instances of complications together without distinction. The main significance of this design is to understand the
tattoo removal process as one that may potentially discourage patients seeking removal from continuing the treatment
process, and cause unnecessary grief
in addition to any preexisting tattoo regret, as is commonly surveyed amongst individuals with tattoos
\cite{serup2017guide}.

Generally, in this paper we apply a full range of inference techniques and machine learning methods to
develop an understanding of best tattoo removal treatment parameters. The
data measures treatment settings used on a tattoo over multiple treatment appointments. Data is provided at both the
patient level and the treatment level for a given tattoo. With regards to methods used, we begin with
analyzing the significance of patient demographic factors that are understood to be correlated with 
complication occurrence by utilizing simple statistical
tests. We then perform parametric and nonparametric tests of significance for
analysis of variance (e.g., Kruskal-Wallis, Wilcoxon rank sum and randomization) and use Tukey's honestly significant
difference (HSD) test to identify important sources of variation. 
Finally, we fit a logistic regression model to identify statistically significant treatment parameters, 
and perform multiple gradient boosting using decision trees to form a ranking of treatment
parameters that a health care practitioner should be most aware of during laser-assisted tattoo removal procedures.

This paper is organized as follows: Section 2 details the dataset and provides descriptions of the methods utilized; Section 3 presents results of the
methods employed; we conclude the paper in Section 4 with last remarks.

\section{Data and Methods}
The individuals that 
come through Homeboy industries are often former gang members, and many of them have tattoos that
may be offensive or attract negative attention. In this section, we discuss the various datasets recorded by Homeboy Industries.
The final cleaned datasets that were used in our analysis detailed tattoo level data and patient demographic data. 

We also utilized many different statistical tests and methods, both parametric and nonparametric. We detail the properties of
these tests and their requirements for legitimate application to the data. 
All computations were done with R-4.0.3 \cite{team2013r}. We used the \texttt{caret} library as the primary driver for applying the
machine learning algorithms in our study. All computing was done on a Linux machine running Ubuntu
20.04 with an Intel 10700K processor.

\subsection{Patient Demographic Data}
Including observations with missing values, there are a total of 2,118 tattoo
observations among 502 patients recorded in the data. We briefly detail the various factors that were recorded at the
patient demographic level in \autoref{tab:Table1}. We note the importance of the Fitzpatrick score---a measure developed
in 1975 to classify skin tones \cite{fitzpatrick1975soleil}---as it is currently  
believed in the literature that the efficacy of selected treatment parameters for laser-assisted
tattoo removal is dependent on the Fitzpatrick score. 

    \begin{table}[!ht]
    \centering
    \captionsetup{font=footnotesize}
    \caption{Table detailing patient demographic and patient-level factors of interest.}
    \rowcolors{2}{gray!20}{white}
    \begin{tabular}{ l p{12cm} }
        \hline
        \textbf{Variable}&\textbf{Description}\\
        \hline
        Patient Age & Age of the patient as of 10 June 2020 (integer-valued)\\
        Sex & Male or female (one-hot encoded as male/not-male, binary)\\ 
        Ethnicity &  Hispanic/Latino, or not (one-hot encoded, binary)\\ 
        Race & Pacific Islander, American/Alaskan Indian, Black, Asian,
        Latino/Hispanic, white, multiracial, other (nominal)\\ 
        Treatment Total & Total number of treatment visits at Homeboy for tattoo removal (integer-valued)\\
        Total Tattoos & Total number of tattoos listed by patient (integer-valued)\\ 
        Fitzpatrick Score & Fairest tone (I) to deeply pigmented (VI) (ordinal)\\
        Complications & Indication of whether a patient ever experienced any complication (one-hot encoded, binary)\\
        \hline
    \end{tabular}
    \label{tab:Table1}
\end{table} 

\subsection{Tattoo Level Data}
At the tattoo level, the data provided by Homeboy Industries includes characteristics of tattoos that went through laser
assisted tattoo removal. At the tattoo level we perform statistical analysis on various factors grouped by
whether the observed tattoo ever experienced a complication in the treatment sequence. 
We briefly detail the tattoo level features in \autoref{tab:Table2}. The data at the tattoo level are important because
it provides insight into the significance of tattoo composition in whether a complication is likely to develop as a
result of undergoing laser treatment as recorded, as well as challenge some assumptions of existing best practices. 

\begin{table}[!ht]
    \centering
    \captionsetup{font=footnotesize}
    \caption{Table detailing tattoo level factors of interest.}
    \rowcolors{2}{gray!20}{white}
    \begin{tabular}{ l p{12cm} }
        \hline
        \textbf{Variable}&\textbf{Description}\\
        \hline
        Category & Tattoo location on body (nominal)\\ 
        Age & Age of tattoo (integer-valued)\\
        Colors & Whether the tattoo was black/blue or had other colors (one-hot encoded as black-blue/not-black-blue,
        binary)\\ 
        Professional & Whether the tattoo was done professionally, or by an amateur (one-hot encoded as
        professional/not-professional, binary)\\ 
        Treatment Total & Total number of treatments recorded for a tattoo (integer-valued)\\ 
        Fitzpatrick Score & Fitzpatrick skin tone rate: fairest tone at I to deeply pigmented at VI (ordinal)\\
        Complications & Indication of whether the tattoo ever experienced any complication at all (one-hot encoded, binary)\\
        \hline
    \end{tabular}
    \label{tab:Table2}
\end{table}


\subsection{Treatment Level Data}
For each tattoo at the treatment level we detail four main parameters involved in laser-assisted tattoo removal,
parameters that are actual tattoo-removal laser settings: fluence, spot size, wavelength and
frequency. We briefly detail the treatment-level features in \autoref{tab:Table3}. Apart from observing complication
occurrence/non-occurrence as a response variable in our analysis with respect to tattoo-level factors and patient
demographic features, we also study the variation of treatment parameters (e.g., fluence, spot size) selected by 
practitioners given a particular tattoo characteristic such as tattoo color and age. The intention for viewing treatment
parameters as a response characteristic is to gauge whether current clinician practices are at all particular to tattoo-level
characteristics at a statistically significant level. 

In our study, some additional design choices were made with regards to studying the treatment parameters over a given
treatment sequence. Since the composition of tattoos have great variation
between and within different tattoo characteristics (e.g., color, ink composition and size) the number of
treatments recorded varied greatly between tattoos. For example, a treatment sequence may have
lasted over 10 weeks for one recorded tattoo, another may have had only a single appointment. 
Thus, two variations of variable transformation on the 
laser treatment parameters were made. 

\begin{center}
\begin{figure}[!ht]
    \includegraphics[width=1\textwidth]{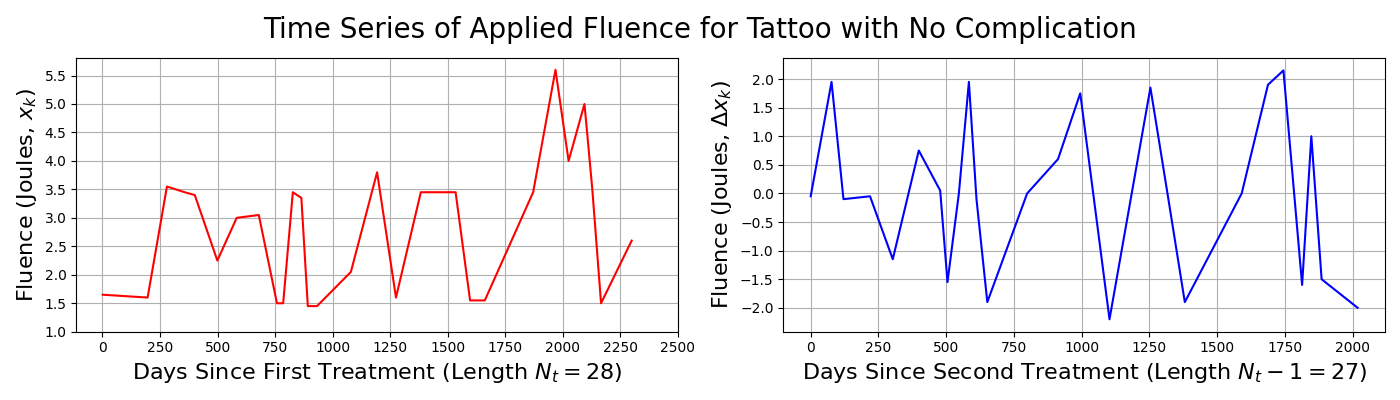}
    \includegraphics[width=1\textwidth]{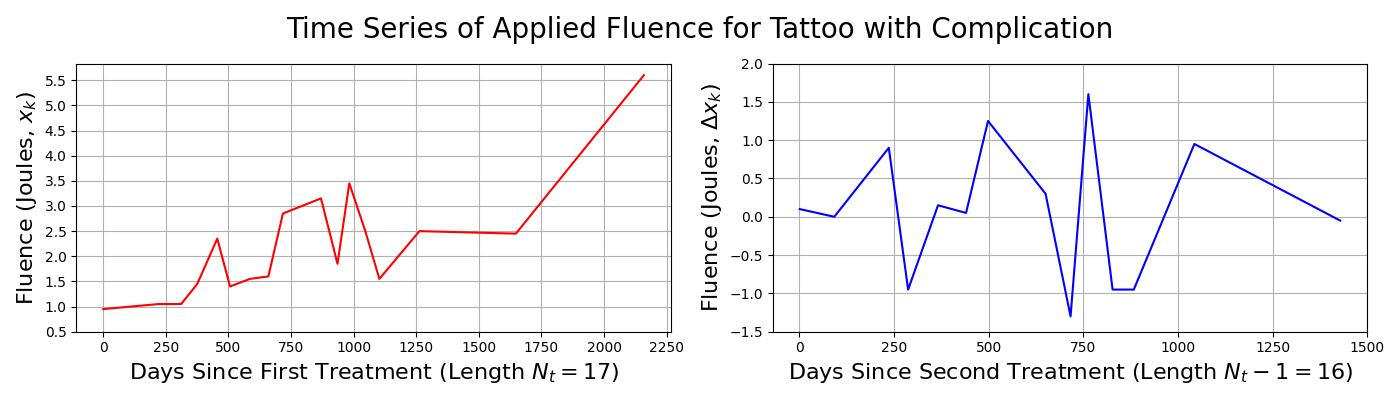}
    \captionsetup{font=footnotesize}
    \caption{We show fluence levels fluctuating by clinician application of laser-assisted tattoo removal. The plots
        in the first row are of fluence settings for a tattoo that does not develop any complications, while the
        plots in the second row
        are of one that has experienced complications. The red time series graphs represents the recorded fluence
        levels, while the blue time series graphs represent the corresponding forward differenced time series of fluence
    levels.}
\label{fig:figure1}
\end{figure}
\end{center}

\begin{table}[!ht]
    \centering
    \captionsetup{font=footnotesize}
    \caption{Table detailing treatment level variables of interest.}
    \rowcolors{2}{gray!20}{white}
    \begin{tabular}{ l p{12cm} }
        \hline
        \textbf{Variable}&\textbf{Description}\\
        \hline
        Fluence & Laser heat and energy intensity measured in joules/cm$^2$ (continuous)\\
        Spot Size & Laser spot radius measured in millimeters (continuous) \\ 
        Wavelength & Laser wavelength at two levels (532nm and 1064nm, binary)\\
        Frequency & Laser frequency measured in Hertz at two levels (5Hz and 10Hz, binary) \\
        Treatment Day & Days since first treatment (integer-valued) \\
        \hline
    \end{tabular}
    \label{tab:Table3}
\end{table} 
The first variation is to simply record the mean and standard
deviation of the above treatment settings for the entire period in which a tattoo had undergone treatment; this
variation was used to study current clinician practices at Homeboy Industries to see how responses of laser settings
changed (i.e., how clinicians selected different treatment parameters) 
on average as a result of varying tattoo characteristics. The second variation also records the means and
standard deviations of the distribution of treatment parameters in sequence, but only up 
to the first instance of a complication occurring; naturally, if a
complication never occurs in a tattoo, the mean and standard deviation of applied laser settings are computed over the entire treatment period.
\autoref{fig:figure1} shows time series of fluence (in red), one of the four possible laser parameters, over different tattoos
with data from the full time series up until the first-arrival of a complication---as well as change the of laser
fluence between treatments (in blue), to be detailed below. Note the irregular duration of
time intervals between treatments, as well as clear nonlinear
fluctuations of fluence chosen by clinicians. 
\vspace{-.8cm}

Furthermore, as an extension of the first complication arrival data, we create additional variables of means and
standard deviations with forward finite differences. For a given variable, representing a time-series treatment
parameter, we compute mean and standard deviations for
\[\Delta x_k =x_{k}-x_{k-1},\]
where the terminating point in the time series, $x_t$, is dependent on whether we are observing first-arrivals of
complications or the entire treatment sequence for a treated tattoo. For example, if we are observing laser fluence for first-arrivals data we produce a forward finite difference variable
$\Delta x$, where the time series terminates at the index in the sequence corresponding to the treatment where a
complication was first reported. Consequently, when computing the mean and standard deviations with respect to this
forward finite difference our expressions simplify as
\[\overline{\Delta x}=\frac{1}{N_t-1}\left(\sum_{k=2}^{N_t}\Delta
x_{k}\right)=\frac{1}{N_t-1}\sum_{k=2}^{N_t}\left(x_{k}-x_{k-1}\right)=\frac{1}{N_t-1}\left(x_{N_t}-x_1\right),\]
where we are essentially measuring the average likelihood of a given treatment parameter setting from the first appointment to the
terminal appointment (i.e., the full treatment sequence for original treatment-level dataset in its entirety, and the truncated treatment sequence for the
first-arrivals dataset). Similarly, for standard deviations of the differenced time series:
\[
    \hat\sigma_{\Delta x}=\sqrt{\text{Var}{\left(\Delta x\right)}}=\frac{1}{N_t-1}\sum_{k=2}^{N_t}\left(\Delta
    x_{k}-\overline{\Delta x}\right)^2=\frac{1}{N_t-1}\left(\sum_{k=2}^{N_t}(\Delta x_k)^2\right)-\left(\overline{\Delta
x}\right)^2,
\]
where $N_t$ is the number of treatments given to a particular tattoo. 
The purpose of this variable differencing is to eliminate any apparent trend; a first order forward differencing
effectively applies a linear filter to our time series \cite{shumway2000time}. The plots of the
time series in \autoref{fig:figure1} demonstrate the variation of $x_k$ and $\Delta x_k$ with respect to laser fluence. 
Tattoo removal clinicians generally aim to
gradually increase the fluence over the treatment sequence, the effect of this difference is to essentially flatten this
linearly trending process, and characterize the distribution of applied laser treatment settings into statistical
parameters (e.g., mean and standard deviation).

\subsection{Statistical Inference} 

We applied both parametric and nonparametric statistical tests of significance to identify statistically significant
factors correlated to complication occurrence in our data (in all our tests we use $\alpha=0.1$). 
We were interested in observing differences in proportions across different patient demographics
and tattoo level factors. We applied the $z$-test for two sample proportion testing as well as the chi-square test of
independence for more than two sample proportions. The hypothesis statement generally for $k$-sample proportions is
\begin{equation*}
    \begin{aligned}
        H_0:p_a-p_b=0 \\
        H_a:p_a-p_b\neq 0
    \end{aligned}
\end{equation*}
where $p_a$ and $p_b$ represent distinct sample proportions, pairwise, across $k$-samples. The necessary assumptions for
these tests were found to be met before their application: independent samples and observations, sufficient sample sizes, 
and mutually exclusive categories in variables \cite{mchugh2013chi}. 

Along with these statistical descriptions of proportions, we additionally were interested in observing current
distribution differences in removal practices and laser treatment settings. We use Welch's two-sample t-test to observe parametric
differences of averages between different treatment settings by whether a tattoo experienced a complication or not
\cite{walpole1993probability, welch1951comparison}. When
the necessary assumptions for a t-test failed---observation independence, sufficient sample size and normality---we
resorted to using the Wilcoxon rank sum test and the randomization test. We strictly applied the Wilcoxon rank sum test
to responses that were categorical-ordinal on independent observations \cite{mann1947test}. The null hypothesis for the
Wilcoxon rank sum test is that the difference in central tendency of the distributions in our data is randomly
occurring, with the alternative being that difference in central tendency is not random, at a statistically
significant level. 

In our implementation of the
randomization test we were interested to see if the measured difference of sample averages for settings applied to
tattoos that developed complications versus those that did not were of random chance. 
The algorithm for the randomization test is as follows:
\begin{enumerate}
    \item Compute and store the true sample average difference. 
        \vspace{-0.3cm}
    \item Randomization process:
        \vspace{-0.2cm}
        \begin{enumerate}
            \item Remove labels of sample designation corresponding to each value; randomly permute data. This step
                corresponds to null hypothesis of observation classes as random.\vspace{-0.1cm}
            \item Relabel observations, partitioning data proportionally into new classes, and measuring the
                differences of the average of the permuted data.
        \vspace{-0.1cm}
            \item Store value as part of an empirical distribution of randomized sample statistics.
        \vspace{-0.1cm}
            \item Iterate for a sufficiently large $N$.
        \end{enumerate}
        \vspace{-0.3cm}
    \item Compute the empirical proportion from the randomization procedure of differences of averages that are greater
        than or less than the true difference (two-tailed variation) for an empirical $p$-value.
\end{enumerate}
The null hypothesis of the randomization test is that the statistic of interest is produced by random
chance, with the alternative being that the observed sample statistic is not randomly produced at a statistically
significant level. Both the Wilcoxon rank sum and randomization tests allow for an intepretation of differences between factors that are
correlated to complication rates, albeit with less statistical power being nonparametric with relative sample
sizes between complication and noncomplication tattoos being heavily unbalanced, 2,000 to 118, respectively \cite{good2013permutation}.

Finally, in observing any true variation across factors with multiple levels we applied one-way analysis of variance
(ANOVA) where parametric assumptions for normal probablity distributions were satisfied, and a Kruskal-Wallis test for
one-way ANOVA where these assumptions were violated. We verified the assumptions of normality, linearity and
homoscedasticity \cite{cox2006principles, douglas2001design}. The hypotheses for one-way ANOVA are
\begin{equation*}
    \begin{aligned}
        H_0&:\mu_1=\mu_2=\cdots=\mu_k\\
        H_a&:\mu_i\neq\mu_j,\;\text{for some}\;i\neq j
    \end{aligned}
\end{equation*}
where the null is that the difference of averages between different categories are not statistically significant, with
the alternative being that there exists a pair samples ($i$ and $j$) in the data where a statistically significant source of variation
in the data is present. In order to locate the source of variation where shown to be significant, we applied Tukey's
HSD test as a multiple comparison among the multi-level factors for which one-way ANOVA
was applied \cite{tukey1949comparing}. In particular, we used the Tukey-Kramer HSD test to account for unequal samples
when computing the \textit{studentized range distribution statistic} for the corresponding confidence interval
\[\bar y_{i\cdot}-\bar y_{j\cdot}\pm \frac{q_{\alpha;k;N-k}}{\sqrt 2}\hat\sigma_\epsilon\sqrt{\frac{1}{n_i}+\frac{1}{n_j}}\frac{}{}\]
where $i$ and $j$ represent distinct sample distributions from different populations; $q_{\alpha;k;N-k}$ follows a studentized range distribution with 
parameters $\alpha$, representing the type II
error; $k$ is the number of populations or levels in a factor; and $N-k$ is degrees of freedom with $N$ being
total number of observations; and without loss of generality, $\bar y_{j\cdot}$ and $n_j$ represents the sample average
for the feature of interest and sample size, respectively \cite{kramer1956extension}. Additionally, 
$\hat\sigma_\epsilon$ denotes the square root of the overall variation of the feature of interest in our total population.

When our parametric conditions failed (e.g., sample distribution normality and homoscedasticity), we resorted to
applying the Kruskal-Wallis test for one-way nonparametric ANOVA \cite{kruskal1952use}. 
The null hypothesis of this test is that the medians
of our populations are equal, and the alternative is that the there is at least one source of random variation among
our populations with a median that is statistically significant among some pair of populations (e.g., median
fluence of settings applied to tattoos on the face versus tattoos on the neck or those on the upper extremeties).

\subsection{Statistical Learning}

In addition to performing inference, we were interested in developing statistical models of our data as related to
complication occurrence (and nonoccurrence) across the first-arrivals tattoo-level dataset.
In our data for complication occurrence, the actual response corresponds to a binary classification problem where
complication occurrence in a given tattoo treatment sequence corresponds to a ``true" outcome and complication
nonoccurrence
corresponds to a ``false" outcome. Generally, logistic regression models a probabilistic relationship of the occurrence
of some response along with its predictors \cite{james2013introduction}. More precisely, logistic regression models the
following:
\[p(X_1,X_2,\hdots,X_n)=P(Y=1|X_1,X_2,\hdots,X_n),\]
where in our case we have a binary label response of complication one-hot encoded as ``1'' and complication
nonoccurrence encoded as ``0'';
and there are $n=10$ predictors, after heuristically removing highly correlated variables using an absolute
correlation threshold of 0.6. The $p$ represents the probability that a given observation with its corresponding
inputs are of an outcome relating to tattoo complication. The precise formulation of the logistic regression model is
\[p(X_1,X_2,\hdots,X_n)=\frac{e^{\beta_0+\beta_1X_1+\cdots+\beta_nX_n}}{1+e^{\beta_0+\beta_1X_1+\cdots+\beta_nX_n}}\]
The coefficients of the logistic regression model are estimated using 
maximimum likelihood estimation (MLE) of the log-likelihood function \cite{czepiel2002maximum}. These coefficients asymptotically follow a
normal distribution, from which we can infer statistical significance and evident dependence on the corresponding
variables to the response in the data. We use this process to identify statistically significant predictors of
treatment parameters corresponding to complications (again, with an $\alpha=0.1$).

In addition to modelling our data to find statistically significant factors corresponding to complication occurrence, 
we wanted to produce a ranking of statistically significant variables in our data. We do this using decision trees,
gradient boosting and bootstrapping \cite{hastie2009elements,efron1994introduction}. Decision trees are machine
learning models that recursively
stratify (split) a feature space---tattoo treatment parameters in our case---into regions that are estimated to be
numerically associated to a given response variable. Boosting is a
general statistical/machine learning paradigm that involves bootstrapping a datatset into many copies,
and sequentially fitting multiple machine learning models to the dataset. 
In this boosting procedure on decision trees there are three
tuning parameters: $B$, the number of trees; $\lambda$, a shrinkage parameter controlling the learning rate of the
boosting method; and $d$, the number of splits in a decision tree, or variables to consider in the feature space
stratification process. 

Boosting with decision trees on a given dataset can produce ranks of variable importance through
the mean decrease in Gini index, a measure of total variance across $K$ classes, formulated as
\[G=\sum_{k=1}^K\hat p_{mk}(1-\hat p_{mk}),\]
where $\hat p_{mk}$ represents the proportion of training observations in the $m$th stratified region corresponding to
the labels of class $k$. The rank of variable importance is measured by the mean decrease in the Gini index over all the
decision trees fitted in the boosting procedure, where the higher the mean decrease in the Gini index for that variable,
the more important it is. From this procedure we additionally bootstrap our dataset
by fitting these gradient boosted trees in our dataset 300 times, with each fit utilizing tuned hyperparameters from repeated
$k$-fold cross validation (8 folds for 2 repeats). We then produce an empirical distribution of ranks of variable
importance for features consdiered in our model; the most frequent rank of a particular feature is then assigned as the
feature's empirical importance rank. The final rankings of these features then decide which features tattoo removal
practitioners should be most aware of in the tattoo removal treatment process. 

\section{Results}
We tabulate our results of statistical inference as well as present the results of our machine learning models.
Throughout we discuss results and the particular application of our methods to data.
\subsection{Statistical Descriptions}
Below are tables corresponding to two-sample (\autoref{tab:Table4}) and $k$-sample proportion tests
(\autoref{tab:Table5}). We measure complication rates across different factors at the patient demographic
and tattoo levels. In our tables, ``Tattoo Complication" refers to whether a single tattoo ever experiences a
complication (i.e., complication occurrence/non-occurrence) within a treatment sequence spanning multiple days. 
\begin{table}[!ht]
    \centering
    \captionsetup{font=footnotesize}
    \caption{Details of two-sample proportions and p-values for $z$-test. Statistically significant factors 
        determining significant proportion differences in complication rates differences are marked by the asterisk.
    The ``True" column represents proportions that are characterized by the factor of interest; ``False" column represents
proportions that do not have the given factor characteristic.}
    \rowcolors{2}{gray!20}{white}
    \begin{tabular}{ p{7.2cm} | l l l}
        \hline
        \textbf{Response Proportion \textit{by} Factor}&\textbf{True}&\textbf{False}&\textbf{$p$-value}\\
        \hline
    Treatment Completion \textbf{by} Complication&       0.14407& 0.07300 &0.0083*\\
        Tattoo Complication \textbf{by} Colored Tattoo&    0.10857&   0.04996&0.0035*\\
    Tattoo Complication \textbf{by} Professional&      0.09351&   0.04291&0.0010*\\
    Complication \textbf{by} Sex (Male/Female) &       0.09964&      0.11764                     &0.6163\\
    Complication \textbf{by} Patient Median Age &0.13061&0.08560 &0.1381\\
    Complication \textbf{by} Patient Ethnicity& 0.12010& 0.06723&0.1452\\
         Complication \textbf{by} Tattoo Median Age              & 0.06379& 0.08560& 0.5562\\
         Complication \textbf{by} Patient Fitzpatrick Score&     0.12000   & 0.10287&   0.5562\\
        \hline
    \end{tabular}
    \label{tab:Table4}
\end{table} 
\newcolumntype{g}{>{\columncolor{gray!20}}l}
\newcolumntype{h}{>{\columncolor{gray!20}}c}
\begin{table}[!ht]
    \centering
    \captionsetup{font=footnotesize}
    \caption{Details of $k$-sample proportions and p-values for chi-squared tests of
    independence. Statistically significant factors determining significant proportion differences in complication rates
are marked by the asterisk.}
        \begin{tabular}{ p{3.7cm} | h | g g g g }
            \rowcolor{white}
        \hline
        \textbf{Reponse \textit{by} Factor} & Parameter & 1st Quartile & 2nd Quartile &       
            3rd Quartile & 4th Quartile \\
        \hline
            \rowcolor{white}
            &Quartile &  4  & 7   &  12  &65\\
            &$n_i$&  122  &  116  &  117  &122\\
            \rowcolor{white}
            \multirow{-3}{10em}{Complication Rate \textbf{by}\\ Total \# of Treatments Tattoo ($p$=6.51e-10*)}
            &$\hat p_i$  & 0.02417& 0.03349&  0.05181&0.12277\\

        \hline
            &Quartile &  4  & 7   & 12   &100\\
            \rowcolor{white}
            \multirow{-2}{10em}{Complication Rate \textbf{by} Total \# of Tattoos per Patient ($p$=0.08395*)}
            &$n_i$  &  92  &  70  & 87   &100\\
            &$\hat p_i$&  0.11957&  0.18571&   0.08046&0.07000\\
            \rowcolor{white}

        \hline
            &Quartile  &  11  & 24   &  52  &72\\
            &$n_i$& 122 & 116 & 117 & 122  \\
            \rowcolor{white}
            \multirow{-3}{10em}{Complication Rate \textbf{by} Total \# of Treatments Patient ($p$=1.35e-06*)}
            &$\hat p_i$&  0.04918&   0.06034&  0.09402&0.24590\\

        \hline
            &Quartile  &   1.4375&  1.9134 & 2.5445   &4.5773 \\
            \rowcolor{white}
            \multirow{-2}{10em}{Complication Rate \textbf{by} Mean Fluence (J/cm$^2$) ($p$=3.532e-05*)}
            &$n_i$  & 508   &   505 &   509 &509\\
            &$\hat p_i$& 0.01771& 0.04554&   0.07466&0.07662\\

        \hline
            \rowcolor{white}
            &Quartile  & 3 yrs   & 7 yrs   & 12 yrs   &40 yrs\\
            &$n_i$ &  292  &306    &233    &253\\
            \rowcolor{white}
        \multirow{-3}{10em}{Tattoo Complication Rate \textbf{by} Tattoo Age ($p$=0.1316)}
            &$\hat p_i$& 0.03425& 0.071895& 0.05150& 0.075099\\

        \hline
            &Quartile  &  29 yrs&  34 yrs  &  41 yrs &80 yrs\\
            \rowcolor{white}
            &$n_i$  & 148   & 109   & 128   &117\\
            \multirow{-3}{10em}{Patient Complication Rate \textbf{by} Patient Age ($p$=0.2807)}
            &$\hat p_i$& 0.09459& 0.07339& 0.14844&0.11111\\
        \hline
    \end{tabular}
    \label{tab:Table5}
\end{table} 

In our application of Welch's $t$-test, Wilcoxon's rank sum and randomization tests, we observe the distribution
differences of settings applied to the tattoos over treatments up until the first arrival of a complication in the
treatment sequence; if a tattoo never experiences a complication the statistics of treatment parameters is computed
across all treatments for that tattoo. \autoref{tab:Table6} shows the results from our t-test as applied to
distributions of time-series statistics (mean and standard deviation) computed over our settings that satisfies some
heuristic observations of normality; \autoref{tab:Table7} shows the results of Wilcoxon rank sum and randomization tests over
distributions of sample statistics that did not satisfy normality and follower highly irregular distribution
shapes (i.e., multimodal and skewed). 
\begin{table}[!ht]
    \centering
    \captionsetup{font=footnotesize}
    \caption{We present the sample averages of these settings along with the corresponding $p$-values from Welch's
    $t$-test. We find no statistically significant treatment settings among these parameters.}
    \rowcolors{2}{gray!20}{white}
    \begin{tabular}{ p{5.3cm} | l l l }
        \hline
        \textbf{Treatment Parameter} & \textbf{Complication} & \textbf{No Complication} & \textbf{$p$-value}   \\
        \hline
        Mean Fluence &  1.958739    &     2.013944   &    0.5713  \\
        Mean Differenced Fluence &0.1848074  &   0.1729800 &    0.8181  \\
        Mean Spot Size& 4.875772 &    4.942978  &   0.4402  \\
        Mean Differenced Spot Size& -0.06056931 &   -0.08294852 &  0.6542    \\
        \hline
    \end{tabular}
    \label{tab:Table6}
\end{table}

\begin{table}[!ht]
    \centering
    \captionsetup{font=footnotesize}
    \caption{We observe nonparameteric differences of statistics of treatment settings that are irregularly distributed
        with respect to the ``first-arrivals" of complications data. 
        For the Wilcoxon rank sum test we compute the sample median, and for the randomization test we compute the
    sample average. For each treatment parameter we present the results in two rows for sample median ($\hat\eta$, Wilcoxon rank sum) 
        and sample mean ($\hat\mu$, randomization).}
    \vspace{3mm}
    \begin{tabular}{ p{4.3cm}| h | g g g }
        \hline
        \rowcolor{white}
        \textbf{Treatment Parameter}& \textbf{Statistic} 
                                    & \textbf{Complication} & \textbf{No Complication} & \textbf{$p$-value}   \\
        \hline
\rowcolor{white}
                    &   $\hat\eta$   &  1064   & 1064   & 9.197e-06*\\
        \multirow{-2}{10em}{Mean Wavelength}&   $\hat\mu$&1010.425   &     1050.169     &7e-04*\\
        \hline
\rowcolor{white}
                &  $\hat\eta$    &   0  & 0   &0.05684*\\   
        \multirow{-2}{10em}{Mean Differenced Frequency}  &  $\hat\mu$    &    0.23336582 &  0.03502956  &1\\
        \hline
\rowcolor{white}&  $\hat\eta$    &  0.80475& 0.72648&0.05307*\\
                  
        \multirow{-2}{10em}{Spot Size Standard Deviation}  &   $\hat\mu$   &   0.8165863  &  0.6899222  &1\\
        \hline
\rowcolor{white}
                    &  $\hat\eta$    &   0  & 0   &1.784e-06*\\
        \multirow{-2}{10em}{Mean Differenced Wavelength}&   $\hat\mu$& 4.663210   &  1.571152   &1\\
        \hline
\rowcolor{white}& $\hat\eta$    &   9.375 & 9.375&0.5299\\

        \multirow{-2}{10em}{Mean Frequency}      & $\hat\mu$     &   9.056429  &  9.145180  &0.5633\\
        \hline
\rowcolor{white}
                    &  $\hat\eta$    &   61  & 71.5   &0.5928\\
        \multirow{-2}{10em}{Average Days Between Treatments}&  $\hat\mu$    &      95.78247&102.58342   &0.6128\\
        \hline
    \end{tabular}
    \label{tab:Table7}
\end{table} 

Additionally, we have our results for current clinician practices and approaches. \autoref{tab:Table8} presents our results
from the Kruskal-Wallis one-way ANOVA for treatment parameters with distributions that violated the 
assumptions required for parametric one-way ANOVA. We applied non-parametric one-way ANOVA to study current 
variations of treatment parameters that were are found to be of practical interest in the literature. In 
particular we were interested in how clinicians applied different settings based on the
tattoo age quartile. 
\begin{table}[!ht]
    \centering
    \captionsetup{font=footnotesize}
    \caption{Sample averages by tattoo age quartile and Kruskal-Wallis $p$-values. All parameters listed were found to
    be statistically signficant (i.e., clinicians exercised distinct application of laser based on tattoo age).}
\rowcolors{2}{gray!20}{white}
\begin{tabular}{p{4.3cm}|cccc|c}
    \textbf{Quartile} & \textbf{1st}(3yrs) & \textbf{2nd} (7yrs) & \textbf{3rd} (12yrs) & \textbf{4th} (40yrs)
                       &\\
\hline
    $n_i$& 292    &306     &233   & 253  &\textbf{$p$-value} \\
\hline
     Mean Fluence ($\hat \mu_i$) & 1.96327   & 2.031156   & 2.019911   & 2.194109 & 0.001946*\\
     Mean Spot Size ($\hat \mu_i$) & 4.888663   &4.849143   & 4.857013   &  4.884559 & 7.314e-05*\\
SD Differ. Frequency ($\hat \mu_i$)  & 2.374471   & 1.948435   & 0.08045977   & 1.948582 & 2.294e-05*\\
\end{tabular}
    \label{tab:Table8}
\end{table}

In our application of parametric one-way ANOVA for studying means, we found particular interest in the application of
average fluence with regards to the location of the tattoo on the body. There indeed exists some statistically significant
variation when with how a clinician chooses a particular fluence level with respect to tattoo location. 
Across tattoo locations on the
body, the distributions of fluence were found to be relatively constant in variation and normally distributed. We
obtained a $p$-value of 3.61-e05 for our one-way ANOVA.
Then, applying Tukey's HSD test among tattoo location, we found that tattoos
applied to the upper extremities and face had the most significant variation in average fluence applied---with sample averages
of 2.09J/cm$^2$ and 1.996J/cm$^2$, respectively---among other
tattoo locations on the body such as lower extremities, back, chest, neck, head and abdomen. 

\subsection{Significant Factors}
We report our machine learning results in this section. \autoref{tab:Table10} shows the numeric results of rank from our
bootstrapped gradient boosting decision trees as well as the coefficient estimates and corresponding $p$-values from our
logistic regression model. It should be noted that for both the logistic regression and gradient-boosted decision tree models,
we removed a few variables from consideration. The removed variables were found to have high absolute correlation
($>0.6$) with other variables in the first-arrival sample statistics dataset of treatment parameters. The intention
of reducing the number of predictors was to simplify our model and reduce any potential inflation of explained variation in our results.

\begin{center}
\begin{figure}[!ht]
\vspace{5mm}
    \includegraphics[width=1\textwidth]{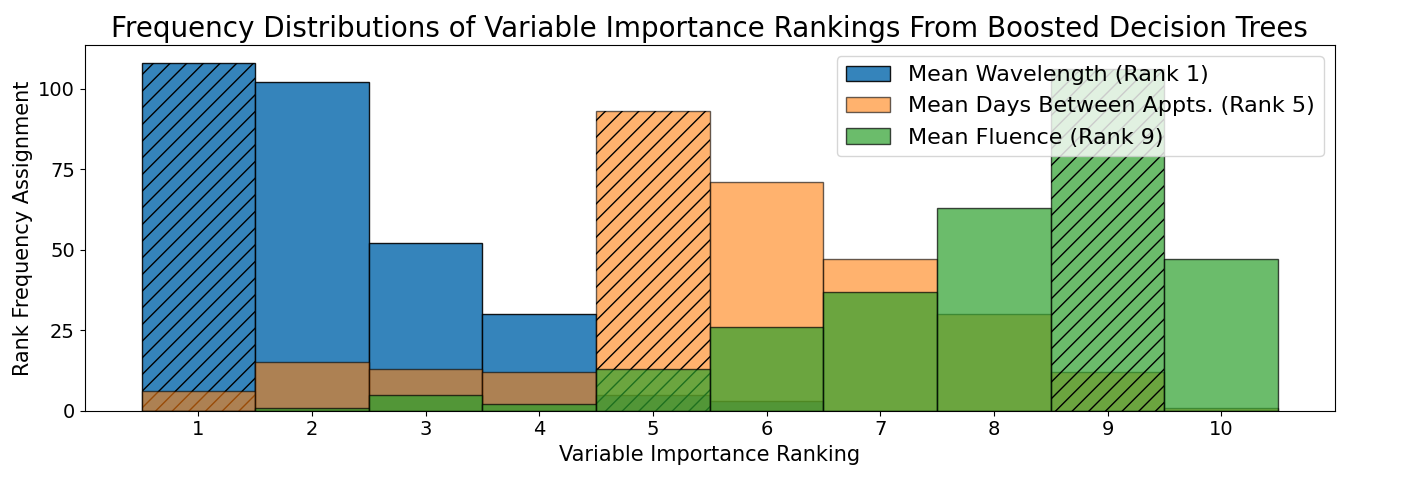}
    \captionsetup{font=footnotesize}
    \caption{Rank distributions of three variables with bars corresponding to ``total rank" hatched. We assign the rankings for each treatment
        parameter based on the mode of their rank. We see that Mean Wavelength (Rank 1), Mean Days Between Appts. (Rank 5), 
        and Mean Fluence (Rank 9) are of descending order of importance
        based on the location of the modes of ranks for each variable from 300 simulations.}
\label{fig:Figure2}
\vspace{-0.3cm}
\end{figure}
\end{center}

With regards to the variable importance ranking system, with over 300 simulations of fitting gradient boosted trees, we 
produced a discrete distribution of ranks for our variables, as detailed in Section 2.5. In cases where the mode of ranks were tied between
two factors, these ties were broken by observing which rank mode was greater. For example,
in \autoref{tab:Table10}, we found that mean wavelength and mean differenced wavelength had rank modes of 1; however,
since the frequency of the rank mode for mean wavelength (100) was greater than the rank mode of the mean differenced
wavelength (90), we reassign ``total ranks" for mean
wavelength as greater than the mean differenced wavelength. \autoref{fig:Figure2} displays a few rank distributions of variable
importance from our simulations. 

\begin{table}[!ht]
    \centering
    \captionsetup{font=footnotesize}
    \caption{Results from boosted decision tree simulations, and coefficient estimates and
    $p$-values for logistic regression coefficients. Statistically significant logistic regression coefficients are
    asterisked. For our
    boosting procedure we show the ranking of important features corresponding to the mode of the variables' frequency
    distribution. Ties in mode frequency rank were broken by comparing relative frequencies. Statistically
    significant logistic regression coefficients are asterisked.}
    \rowcolors{2}{gray!20}{white}
    \begin{tabular}{ p{5.3cm} | h | g | g g }
        \rowcolor{white}
        \hline
        \textbf{Treatment Parameter} & \textbf{Total Rank} & \textbf{Rank Mode} & \textbf{Estimate} &\textbf{$p$-value}   \\
        \hline
        Mean Wavelength         &  1    &    108 (1)    &   -0.004299 & 0.00118*\\
        Mean Differenced Fluence&   2   &  84 (1 $\to$ 2)    &   0.550567    & 0.33951\\
        Mean Differenced Wavelength&  3    &  80 (1 $\to$ 2 $\to$ 3) & 0.004979    & 0.35263\\
        SD Spot Size            &   4   &  117 (4)   &  1.294140  &0.00609* \\
        Mean Days Between Appts.&   5   &    93 (5)   & -0.001966   & 0.35287\\
        Mean Differenced Spot Size&   6   &   76 (5 $\to$ 6)    & 0.449331 &0.38701 \\
        Mean Differenced Frequency&   7   & 64 (6 $\to$ 7)    &    0.466965 &0.04217* \\
        Mean Spot Size          &    8  &     75 (7 $\to$ 8)   &    -0.154691 &0.71596\\
        Mean Fluence            &    9  &   106 (9)    &   -0.137015  &0.59433\\
        Mean Frequency          &  10    &   197 (10)     &   0.014290  &0.93119 \\
        \hline
    \end{tabular}
    \label{tab:Table10}
\end{table} 

\vspace{-12mm}
\section{Conclusion}

For our tattoo data we applied a whole range of methods from classical statistical parametric tests to
modern machine learning algorithms. We characterized the time series treatment data across all parameters using sample
means and standard deviations, as well as having performed forward finite differencing to characterize the variation in
laser-assisted tattoo removal treatment parameters \textit{between} appointments. From the transformations made on our variables, we
applied both parametric and nonparametric tests of significance to identify tattoo-level and
patient demographic characteristics strongly correlated to complication occcurence. 

Below we list response-factor relationships that were found to be statistically significant along with a brief
details with regards to practical implications:
\begin{itemize}
    \item Complication occurrence is related to \textbf{decreased} likelihood of tattoo removal treatment completion.
        Patients may become ambivalent about continuing tattoo treatments if a complication is experienced.
    \item Tattoo removal complications are statistically \textbf{more likely} for tattoos done by a
        professional tattoo artist as opposed to an amateur artist. This may be due to the depth of the tattoo
        placement and/or ink utilized in the professional tattoo process.
    \item Tattoo removal complications are statistically \textbf{more likely} for tattoos with color (e.g., red, green,
        yellow) as opposed to black/blue tattoos. This may be due to the skin reaction sensitivity 
        in tattoos with certain pigments.
    \item Greater total number of treatments on a tattoo, number of tattoos on a patient, and number of treatments a
        patient has undergone are all related to \textbf{increased} complication rates. With regards to tattoo age,
        clinicians apply statistically greater average fluences and have greater variability in spot size applied to
        tattoos over the treatment period. 
    \item Greater average fluence applied to a tattoo is related to \textbf{increased} likelihood of complication
        occurrence.
    \item Greater average laser wavelength over a treatment period is \textit{both} 
        parametrically and non-parametrically related to \textbf{increased}
        likelihood of complication occurrence. Shorter wavelengths may be emulating intense      
        radiation causing complications associated with skin discoloration.
    \item According to our logistic regression model, average wavelength, standard deviation in laser spot size, 
        and overall change of laser frequency applied to a
        tattoo through a given treatment sequence are related to \textbf{increased} likelihood of complication
        occurrence.
\end{itemize}
It is also worth noting the response-factor relationships that were found to \textit{not} be statistically significant
with respect to the current literature (again, with brief explanations and implications):
\begin{itemize}
    \item The average number of days between appointments is not statistically related to complication
        occurrence. The number of days between appointments in the data is sufficiently long for skin recovery
        before a follow-up appointment.
    \item Tattoo age and patient age is not statistically related to the variation in complication rates. This
        possibly suggests that current clinician approaches to tattoo removal take into account tattoo and patient
        age, since the application of certain settings used by clinicians vary between different ages at a
        statistically significant level.
    \item Between groups of tattoos that experienced complications and those that did not, the variation in average laser fluence, overall change in laser fluence between the first and last appointment, average spot
        size, and overall change in laser spot size between the first and last appointment were not found to be
        distinct at a statistically significant level.
    \item Patient sex, ethnicity, Fitzpatrick score, and being above the median age are not statistically
        related to complication occurrence within a recorded treatment sequence, independent of whether a patient
        completed the entire treatment process.
\end{itemize}

Finally, note the discrepancy of our results between the two machine learning models. Although our boosted decision tree
procedure lowly ranks a statistically significant variable---according to our logistic regression model---such as ``Mean
Difference Laser Frequency," this result is not an inconsistent finding. The two models essentially offer distinct
interpretations of the data. 

The logistic regression model measures that in a total interaction of all ten variables, average wavelength, standard
deviation in spot size, and average variation of laser frequency between the first appointment and pre-arrival of a
recorded tattoo complication are statistically significant factors for the tattoo removal practitioner to 
be aware of in the treatment process. On the other hand,
the aggregated boosted decision tree model ranks variables by observing how our predictions of complication
occurrence/non-occurrence are affected if they are not considered in the interaction. For example, excluding average
wavelength in the interaction of treatment parameters will most frequently negatively affect classification of complication
occurrence/nonoccurrence in our data compared to the exclusion of average frequency which will be of least consequence in our
predictions.

Limitations worth noting with respect to our findings are that these results are specific to our given dataset, which has
a fair share of missing values without any data imputation having been performed, and we
do avoid making any full clinical interpretations of our results as well.
Yet, our data is relatively abundant and detailed, and we've applied many interesting techniques along with applying
this ad hoc aggregated variable ranking procedure using gradient boosted decision trees, 
which may prove to be a theoretically well-justified and
statistically powerful way of producing ranks of variable importance.
\subsection*{Acknowledgements}
JH, DN and RY are grateful to Jessica Bogner at Homeboy Industries, Los Angeles, and 
Dr. Jo Marie Reilly at Keck Medicine of USC, for providing data and guidance with this project. The authors were
partially  supported by NSF BIGDATA \#1740325 and NSF DMS \#2011140.
\bibliographystyle{plain}
\bibliography{mybib}

\begin{thebibliography}{10}

\bibitem{baumler2019laser}
W~B{\"a}umler and KT~Wei{\ss}.
\newblock Laser assisted tattoo removal--state of the art and new developments.
\newblock {\em Photochemical \& Photobiological Sciences}, 18(2):349--358,
  2019.

\bibitem{cox2006principles}
David~Roxbee Cox.
\newblock {\em Principles of statistical inference}.
\newblock Cambridge university press, 2006.

\bibitem{czepiel2002maximum}
Scott~A Czepiel.
\newblock Maximum likelihood estimation of logistic regression models: theory
  and implementation.
\newblock {\em Available at czep. net/stat/mlelr. pdf}, pages
  1825252548--1564645290, 2002.

\bibitem{douglas2001design}
C~Montgomery Douglas.
\newblock {\em Design and analysis of experiments}.
\newblock John Wiley and Sons Inc, 2001.

\bibitem{efron1994introduction}
Bradley Efron and Robert~J Tibshirani.
\newblock {\em An introduction to the bootstrap}.
\newblock CRC press, 1994.

\bibitem{fitzpatrick1975soleil}
Thomas~B Fitzpatrick.
\newblock Soleil et peau.
\newblock {\em J Med Esthet}, 2:33--34, 1975.

\bibitem{good2013permutation}
Phillip Good.
\newblock {\em Permutation tests: a practical guide to resampling methods for
  testing hypotheses}.
\newblock Springer Science \& Business Media, 2013.

\bibitem{hastie2009elements}
Trevor Hastie, Robert Tibshirani, and Jerome Friedman.
\newblock {\em The elements of statistical learning: data mining, inference,
  and prediction}.
\newblock Springer Science \& Business Media, 2009.

\bibitem{ho2015laser}
Stephanie~GY Ho and Chee~Leok Goh.
\newblock Laser tattoo removal: a clinical update.
\newblock {\em Journal of cutaneous and aesthetic surgery}, 8(1):9, 2015.

\bibitem{james2013introduction}
Gareth James, Daniela Witten, Trevor Hastie, and Robert Tibshirani.
\newblock {\em An introduction to statistical learning}, volume 112.
\newblock Springer, 2013.

\bibitem{kramer1956extension}
Clyde~Young Kramer.
\newblock Extension of multiple range tests to group means with unequal numbers
  of replications.
\newblock {\em Biometrics}, 12(3):307--310, 1956.

\bibitem{kruskal1952use}
William~H Kruskal and W~Allen Wallis.
\newblock Use of ranks in one-criterion variance analysis.
\newblock {\em Journal of the American statistical Association},
  47(260):583--621, 1952.

\bibitem{kurniadi2020laser}
Ivan Kurniadi, Farida Tabri, Asnawi Madjid, Anis~Irawan Anwar, and Widya
  Widita.
\newblock Laser tattoo removal: Fundamental principles and practical approach.
\newblock {\em Dermatologic Therapy}, page e14418, 2020.

\bibitem{mann1947test}
Henry~B Mann and Donald~R Whitney.
\newblock On a test of whether one of two random variables is stochastically
  larger than the other.
\newblock {\em The annals of mathematical statistics}, pages 50--60, 1947.

\bibitem{mchugh2013chi}
Mary~L McHugh.
\newblock The chi-square test of independence.
\newblock {\em Biochemia medica}, 23(2):143--149, 2013.

\bibitem{serup2017guide}
J{\o}rgen Serup and Wolfgang B{\"a}umler.
\newblock Guide to treatment of tattoo complications and tattoo removal.
\newblock In {\em Diagnosis and Therapy of Tattoo Complications}, volume~52,
  pages 132--138. Karger Publishers, 2017.

\bibitem{shumway2000time}
Robert~H Shumway, David~S Stoffer, and David~S Stoffer.
\newblock {\em Time series analysis and its applications}, volume~3.
\newblock Springer, 2000.

\bibitem{team2013r}
R~Core Team et~al.
\newblock R: A language and environment for statistical computing.
\newblock 2013.

\bibitem{tukey1949comparing}
John~W Tukey.
\newblock Comparing individual means in the analysis of variance.
\newblock {\em Biometrics}, pages 99--114, 1949.

\bibitem{walpole1993probability}
Ronald~E Walpole, Raymond~H Myers, Sharon~L Myers, and Keying Ye.
\newblock {\em Probability and statistics for engineers and scientists},
  volume~5.
\newblock Macmillan New York, 1993.

\bibitem{welch1951comparison}
Bernard~Lewis Welch.
\newblock On the comparison of several mean values: an alternative approach.
\newblock {\em Biometrika}, 38(3/4):330--336, 1951.

\end{thebibliography}
\end{document}